\documentclass[%
 reprint,
 amsmath,
 amssymb,
 aps,
 prl,
floatfix,
groupedaddress,
]{revtex4-2}

\usepackage{graphicx}
\usepackage{dcolumn} 
\usepackage{bm} 
\usepackage{xcolor}
\usepackage{hyperref} 
\usepackage{chemformula}
\usepackage[mathlines]{lineno} 
\usepackage{academicons}
\usepackage{mathtools}

\usepackage{nomencl}

\makenomenclature

\newcommand{\opp}[1]{\operatorname{#1}}
\newcommand{\supp}[1]{\opp{supp}({#1})}

\newcommand{\UCB}{
  Department of Materials Science and Engineering,
  University of California Berkeley,
  Berkeley CA, USA
}
\newcommand{\LBL}{
  Materials Sciences Division,
  Lawrence Berkeley National Laboratory,
  Berkeley CA, USA
}

\begin{document}


\title{The cluster decomposition of the configurational energy of multicomponent alloys}

\author{Luis Barroso-Luque}
\email{lbluque@berkeley.edu}
\affiliation{\LBL}

\author{Gerbrand Ceder}
\email{gceder@berkeley.edu}
\affiliation{\UCB}
\affiliation{\LBL}

\begin{abstract}
Lattice models parameterized using first-principles calculations constitute an effective framework to simulate the thermodynamic behavior of physical systems. The cluster expansion method is a flexible lattice-based method used extensively in the study of multicomponent alloys. Yet despite its prevalent use, a well-defined understanding of expansion terms has remained elusive. In this letter, we introduce the \emph{cluster decomposition} as a unique and basis-agnostic decomposition of any general function of the atomic configuration in a crystal. We demonstrate that cluster expansions constructed from arbitrary orthonormal basis sets are all representations of the same cluster decomposition. We show how the norms of expansion coefficients associated with the same crystallographic orbit are invariant to changes between orthonormal bases. Based on its uniqueness and orthogonality properties, we identify the cluster decomposition as an invariant ANOVA decomposition. We leverage these results to illustrate how functional analysis of variance and sensitivity analysis can be used to directly interpret interactions among species and gain insight into computed thermodynamic properties. The work we present in this letter opens new directions for parameter estimation, interpretation, and use of applied lattice models on well-established mathematical and statistical grounds.
\end{abstract}

\pacs{Valid PACS appear here}

\maketitle

\clearpage
Computational methods based on lattice models are used extensively in the applied physical sciences. Parameterized lattice models are actively used in materials science to study metallic alloys \cite{hart_machine_2021, sutton_first-principles_2020, nataraj_systematic_2021}, semi-conductors \cite{xu_cluster_2019, han_atomistic_2022}, super-ionic conductors \cite{richards_design_2016-1, deng_phase_2020}, battery electrodes \cite{van_der_ven_rechargeable_2020}, and surface catalysis \cite{chen_computational_2021}. The cluster expansion (CE) method constitutes a mathematical formalism for the representation and parameterization of generalized lattice models \cite{sanchez_generalized_1984, barroso-luque_cluster_2022, xie_perspective_2022}. The CE method coupled with Monte Carlo (MC) sampling has become an established technique to compute thermodynamic properties of multi-component crystals \cite{van_de_walle_automating_2002, van_der_ven_first-principles_2018}. Recent advancements have introduced generative models as alternative ways to compute free energies \cite{wu_solving_2019, damewood_sampling_2022}. Additionally, the underlying mathematical structure and formalism of the CE have been used to develop methodological extensions to parameterize functions of continuous degrees of freedom \cite{drautz_spin-cluster_2004, singer_spin_2011, thomas_exploration_2017, thomas_hamiltonians_2018} that can also be used to represent vector and tensor material properties \cite{van_de_walle_complete_2008, drautz_atomic_2020}, and even capture full potential energy landscapes \cite{drautz_atomic_2019}.

The core of the CE method is the expansion of a function of configurational variables distributed on a lattice. The expansion is expressed in terms of \emph{correlation functions}, which are constructed by averaging over functions that act on symmetrically equivalent clusters of sites and so ensure that the symmetries of the physical system are respected. Formally, the mathematical formalism of the CE constitutes a harmonic expansion of functions over a tensor product domain \cite{ceccherini-silberstein_discrete_2018}. Using correlation functions that operate over small subsets of variables permits tractable parameterization and calculations of complex properties.

Intuitively, such a formalism leads to expansions that are generalizations of the Ising model \cite{wolverton_ising-like_1995, brush_history_1967},
\begin{equation}
    H(\bm{\sigma}) = \sum_{\beta}J_\beta \sum_{\alpha\in\beta}\prod_{i\in[N]} \phi_{\alpha_i}(\bm{\sigma}_i),
    \label{eq:ce-expanded}
\end{equation}
where $\bm{\sigma}$ is a string of occupation variables that represent the chemical species residing on each of $N$ sites; $\alpha$ are multi-indices of length equal to $N$; $\beta$ are sets of symmetrically equivalent multi-indices; $J_\beta$ are expansion coefficients. The site functions $\phi_{\alpha_i}$ are taken from basis sets spanning the single variable function space over the corresponding occupation variable $\bm{\sigma}_i$. The product of site functions over all sites is referred to as a product function or a \emph{cluster basis function} \cite{van_der_ven_first-principles_2018}, which we write compactly as $\Phi_{\alpha}$.


The resemblance to the Ising model is evident when considering binary occupation variables; for which the monomials $\phi_0 = 1$ and $\phi_1(\bm{\sigma}_i) = \pm 1$ can be used as a site basis. For the case of an arbitrary number of components $\bm{\sigma}_i \in \Omega_i$, the requirements are simply that the constant function $\phi_0 = 1$ is included and that the basis is orthonormal under the following inner product \cite{sanchez_cluster_1993, sanchez_cluster_2010},
\begin{equation}
    \langle\phi_j, \phi_k \rangle = \sum_{\bm{\sigma}_i\in\Omega_i}\rho_i(\bm{\sigma}_i)\phi_j(\bm{\sigma}_i)\phi_k(\bm{\sigma}_i)
    \label{eq:inner-prod}
\end{equation}
where $\rho_i(\bm{\sigma}_i)$ is an a-priori probability measure over the allowed values of  $\bm{\sigma}_i \in \Omega_i$. The inner product in Equation \ref{eq:inner-prod} can be interpreted as the expected value in the non-interacting limit. A uniform probability measure is most often used, but generally, it can be equal to the concentration of chemical species in the non-interacting limit \cite{sanchez_cluster_2010}. We will call a site basis that satisfies the above two requirements a \emph{standard site basis}.

By including $\phi_0 = 1$ in all site bases, Equation \ref{eq:ce-expanded} is a \emph{hierarchical} expansion, where each function $\Phi_\alpha$ has as an effective domain the occupation variables of a cluster of sites $S$ given by the \emph{support} of its multi-index $\opp{supp}(\alpha)$. Leveraging this hierarchical framework, the cluster functions can be written solely in terms of clusters of sites $S = \opp{supp}(\alpha)$ and the nonzero entries of the multi-indices, which we call \emph{contracted multi-indices} $\widehat{\alpha}$,
\begin{equation}
        \Phi_{\alpha}(\bm{\sigma}) = \Phi_{\widehat{\alpha}}(\bm{\sigma}_S) = \prod_{i=1}^{|S|}\phi_{\widehat{\alpha}_i}(\bm{\sigma}_{S_i})
    \label{eq:cluster-function}
\end{equation}
Expression \ref{eq:cluster-function} makes the effective domain of cluster functions explicit. Additionally, Equation \ref{eq:cluster-function} separates the functional form of a cluster function and the particular cluster of sites it acts on. Meaning that cluster functions that operate on symmetrically equivalent clusters have the same functional form (indicated by $\widehat{\alpha}$), but differ in their effective domain (indicated by $S$). We refer to cluster functions that are constructed using a standard site basis as \emph{Fourier cluster functions}, and a resulting expansion as a \emph{Fourier CE}.

The requirement that site basis functions be orthonormal ensures that the resulting set of cluster functions is itself orthonormal \cite{sanchez_generalized_1984, ceccherini-silberstein_discrete_2018}. However, orthonormality is not a strict requirement, since a set of cluster functions $\Phi_\alpha$ based on any site basis will span the space of functions over configuration \cite{ceccherini-silberstein_discrete_2018}. In fact, there exist many applications of CE methodology that use non-orthogonal basis sets \cite{stampfl_first-principles_1999, drautz_obtaining_2006, zhang_cluster_2016, barroso-luque_sparse_2021, kim_multisublattice_2022}. Insightful connections to renowned classical lattice models exist for both non-orthogonal and Fourier CEs. A binary Fourier CE is a direct generalization of the Ising model to higher-degree interactions. Similarly, a binary CE using indicator  functions $\phi_1 = \bm{1}_{\sigma}$ is a generalization of the \emph{lattice gas} model, or a generalization of the Potts model when an overcomplete frame representation is used \cite{barroso-luque_sparse_2021, kim_multisublattice_2022}. Such connections to classical lattice models have been used by practitioners to evaluate the spatial decay of interactions \cite{sanchez_cluster_2010, pei_statistics_2020} and to analyze the effects of specific species and their interactions on the total energy \cite{deng_phase_2020, kim_multisublattice_2022, xie_perspective_2022} by examining the fitted expansion coefficients. However, for complex systems with three or more components, coefficient values depend non-trivially on the particular choice among numerous possible basis sets,\footnote{The number of distinct basis in lattice gas CE sets grows with the number of components. In a Fourier CE there are infinitely many basis set choices for 3 or more components. In an overcomplete representation \cite{barroso-luque_sparse_2021} there are infinitely many expansion coefficient choices that represent a given Hamiltonian.} and direct interpretation of coefficients leaning on intuition from the Ising or lattice gas models can be precarious and ambiguous.

In this letter, we show that a Fourier CE can be expressed as a unique basis agnostic decomposition which we call the \emph{cluster decomposition}. The cluster decomposition is related to well-established expansions of random variables known as ANOVA or Sobol decompositions \cite{sobol_global_2001} among other names \cite{hoeffding_class_1948, efron_jackknife_1981}. Moreover, the cluster decomposition has analytic properties that lead to a deeper understanding of the structure and interpretation of expansion terms. We then illustrate a practical use case of the cluster decomposition based on related concepts from functional analysis of variance (fANOVA) and sensitivity analysis (SA) as a means to gain mathematically rigorous insight from CE and MC simulations of real materials.

\begin{figure}
    \centering
    \includegraphics[width=0.5\textwidth]{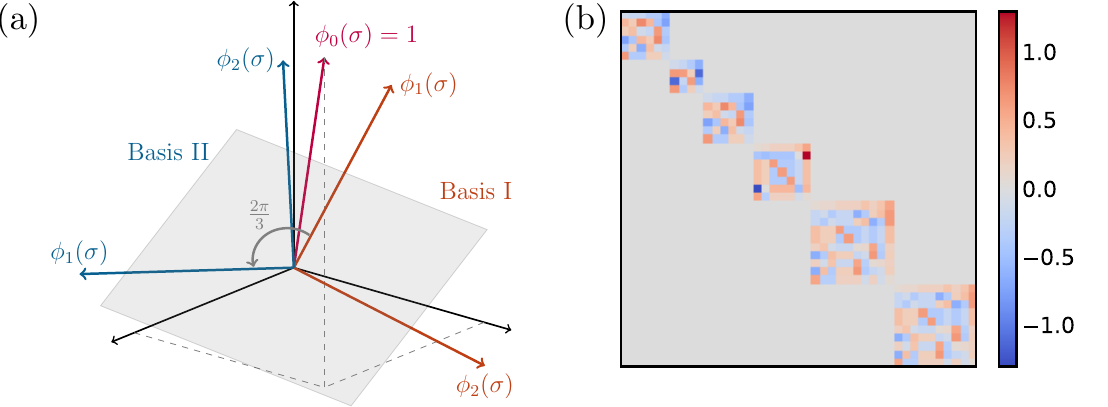}
    \caption{(a) Geometry of standard site basis sets for a ternary site space. Two standard site basis sets related by a rotation of $2\pi/3$ are shown. Both basis sets include the constant $\phi_0$. (b) Change of basis matrix relating the two different sets of Fourier cluster basis functions up to quadruplets constructed using the site basis sets in (a).}
    \label{fig:invariance}
\end{figure}

Let us first motivate the search for a basis-agnostic representation of a CE from a geometric observation. By virtue of their aforementioned properties, it follows that standard site basis sets are related by rotations about the hyperplane normal to the constant function $\phi_0 = 1$. This observation is illustrated graphically for a ternary site space in Figure \ref{fig:invariance}a. Any standard site basis must include two orthogonal basis functions that lie on the plane orthogonal to $\phi_0$. The geometry of standard site bases implies that the change of basis matrix (CBM) $M$ between two resulting Fourier cluster basis sets is given by products of site basis rotations,
\begin{equation}
    M_{\gamma, \alpha} = \left(\prod_i^NR_{\alpha_i,\gamma_i}\right)\delta_{\opp{supp}(\gamma)\opp{supp}(\alpha)}
    \label{eq:cbm}
\end{equation}
where $\gamma$ and $\alpha$ are multi-indices for two Fourier cluster basis sets.

The CBM is block-diagonal---any term connecting cluster functions of symmetrically distinct clusters are zero. Further, since the CBM is also unitary, it follows that the blocks themselves are unitary, implying that the norm of expansion coefficients within each block is conserved. A visualization of the block-diagonal CBM between two Fourier cluster bases of a ternary system including up to quadruplet terms is shown in Figure \ref{fig:invariance}b.

To continue, we define \emph{reduced correlation functions} as the average of cluster functions over symmetrically equivalent contracted multi-indices $\widehat{\alpha}$,
\begin{equation}
    \widehat{\Theta}_{\beta}(\bm{\sigma}_S) = \frac{1}{\widehat{m}_{\beta}} \sum_{\widehat{\alpha}\in\widehat{\beta}}\Phi_{\widehat{\alpha}}(\bm{\sigma}_S),
\end{equation}
where $\widehat{\beta}$ is a set (orbit) of symmetrically equivalent contracted multi-indices $\widehat{\alpha}$, i.e. symmetrically equivalent site function permutations over a fixed cluster of sites $S$. $\widehat{m}_\beta$ is the total number of contracted multi-indices in $\widehat{\beta}$.

Using reduced correlation functions we rewrite Equation \ref{eq:ce-expanded} as follows,
\begin{equation}
    H(\bm{\sigma}) = \sum_B \sum_{\widehat{\beta}\in \widehat{L}(B)} \widehat{m}_{\beta} J_\beta \sum_{S\in B}\widehat{\Theta}_\beta(\bm{\sigma}_S),
    \label{eq:ce-re}
\end{equation}
where $B$ are orbits of symmetrically equivalent clusters of sites $S\subseteq [N]$; and $\widehat{L}(B)$ are sets of orbits of contracted multi-indices $\widehat{\beta}$, which represent symmetrically distinct labelings over the sites in the clusters $S\in B$.


The two inner sums in Equation \ref{eq:ce-re} are independent and can be re-arranged to obtain a far more physically intuitive many-body expansion as follows,
\begin{equation}
    H(\bm{\sigma}) = \sum_B  \sum_{S\in B} \widehat{H}_B(\bm{\sigma}_S)  \\
    \label{eq:decomp-ext}
\end{equation}
where the $n$-body terms $\widehat{H}_B(\bm{\sigma}_S)$ account for the energy originating from the interactions amongst the species residing on the clusters $S\in B$. For clusters $S$ with more than one site, $|S| > 1$, we call these terms \emph{cluster interactions}.

Following the original CE formalism, Equation \ref{eq:decomp-ext} can also be written as a density by using averages of cluster interactions $\widehat{H}_B$ over symmetrically equivalent clusters $S\in B$,
\begin{align}
    H(\bm{\sigma}) &= N \sum_B  m_B \left(\frac{1}{m_B N} \sum_{S\in B} \widehat{H}_B(\bm{\sigma}_S)\right) \nonumber \\
    &= N\sum_B  m_B  H_B(\bm{\sigma}),
    \label{eq:decomp-int}
\end{align}
we will refer to the terms $H_B$ with $|S| > 1$ for all $S\in B$ as \emph{mean cluster interactions}, and as \emph{composition effects} for point clusters ($|S| = 1$).

Equations \ref{eq:decomp-ext} and \ref{eq:decomp-int} are the \emph{cluster decomposition} of the Hamiltonian $H(\bm{\sigma})$. Note that although such an expression can be obtained for any choice of site basis---orthogonal or not---a true cluster decomposition is obtained from a Fourier CE only. This distinction is fundamental since CE expansions using non-orthogonal basis sets will not have the analytical properties that we describe in the remainder of this letter.

It follows directly from our previous analysis of the geometry of Fourier cluster functions, that cluster interactions are invariant to a change of standard basis, i.e. they are invariant to arbitrary rotations orthogonal to $\phi_0$. As a result, the norm of the cluster interactions,
\begin{equation}
    ||\widehat{H}_B||^2_2 = \sum_{\widehat{\beta}\in \widehat{L}(B)} \widehat{m}_{\beta} J^2_\beta
    \label{eq:cluster-weight}
\end{equation}
is invariant to the choice of standard site basis. In line with CE and discrete Fourier expansion terminology, we will call the squared norm of a cluster interaction $||\widehat{H}_B||^2_2$ the \emph{effective cluster weight} of a cluster $S\in B$. In addition, we define the \emph{total cluster weight} as the effective cluster weight multiplied by the multiplicity of its orbit, $m_B ||\widehat{H}_B||^2_2$.

Cluster interactions have the following significant mathematical properties \footnote{Derivations and proofs are given in the Supplemental Material}:
\begin{enumerate}
    \item $\langle H_B \rangle = 0$ (zero mean)
    \item $\langle H_B, H_D \rangle = 0$ for $B\neq D$ (orthogonal)
    \item $\langle H_B, F_{\mathcal{D}} \rangle = 0$ for any set of orbits $\mathcal{D}$ such that $B \notin \mathcal{D}$ and any function $F_{\mathcal{D}}$ that can be expanded using Fourier basis functions $\Phi_\alpha$ with $\opp{supp}(\alpha)\in D$ for $D \in \mathcal{D}$. (irreducible)
\end{enumerate}

From properties (1) and (2) it follows that the cluster decomposition of $H$ is \emph{unique} \cite{hooker_generalized_2007}; meaning there exists one and only one set of cluster interactions $\widehat{H}_B$ for any given Hamiltonian $H$. Equivalently, property (1) implies that Equations \ref{eq:decomp-ext} and \ref{eq:decomp-int} are \emph{ANOVA-representations} of $H(\bm{\sigma})$ \cite{sobol_global_2001, hooker_generalized_2007}. In fact, re-written in such a form, a CE using a standard basis is nothing more than an fANOVA representation, in which by symmetry, interactions among equivalent clusters $S\in B$ are given by the same function $H_B$. By this consideration, using a cluster decomposition as an effective Hamiltonian to define a Boltzmann distribution can be thought of as log-density ANOVA estimation of a probabilistic graphical model \cite{jeon_effective_2006, jeon_characterization_2012, gu_regression_2013-2}.

\begin{figure*}
    \centering
    \includegraphics[width=\textwidth]{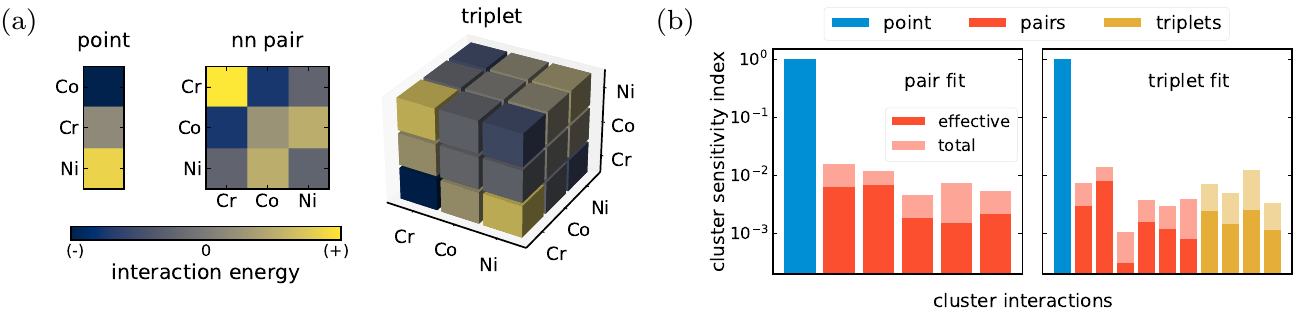}
    \caption{(a) Visualization of the main effect (point), nearest neighbor pair, and triplet cluster interactions as tensors for a cluster decomposition of the configuration energy of a \ch{NiCoCr} alloy using a fit that includes pairs and triplets with diameters up to up to $9 \AA$ and $4.3 \AA$ respectively. (b) Cluster sensitivity indices of two fitted \ch{NiCoCr} cluster decompositions (one including only pairs, and another including pairs and triplets) sorted by cluster diameter. Effective (total) cluster indices are shown with solid (translucent) colors.}
    \label{fig:interactions}
\end{figure*}

Using the construction of ANOVA representations, we can now obtain a much deeper understanding of the terms in a CE. Precisely, ANOVA terms are constructed from hierarchical inclusion-exclusion of means conditioned on the occupancy of clusters. For example, it is already known from the original CE formalism \cite{sanchez_generalized_1984} that the constant term is equal to the mean of the Hamiltonian. $J_{\emptyset} = H_{\emptyset} = \langle H(\bm{\sigma}) \rangle$. In the statistics literature, $J_{\emptyset}$ is usually referred to as the \emph{grand mean} \cite{gelman_analysis_2005}. The single site terms terms, $\widehat{H}_{P}(\bm{\sigma}_i)$ are the difference between the mean conditioned on the $i$-th site and the grand mean, $\widehat{H}_{P}(\bm{\sigma}_i) = \langle H(\bm{\sigma})\;\rvert\;\bm{\sigma}_i\rangle - \langle H(\bm{\sigma})\rangle$. The point terms of an ANOVA representation are called \emph{main effects} \cite{gelman_analysis_2005}. The main effects are the mean contribution that a specific species $\bm{\sigma}_i$ residing on the $i$-th site has on the total energy. The average of \emph{main effects} in the cluster decomposition (a term $H_P$ in Equation \ref{eq:decomp-int}) represents the portion of the Hamiltonian that depends on composition only.

The remaining terms involving clusters $S$ with more than one site are known as \emph{interactions} \cite{gelman_analysis_2005}, motivating our terminology. A cluster interaction $\widehat{H}_{B}(\bm{\sigma}_S)$ of cluster $S$ is computed as the mean conditioned on the sites in cluster $S$, minus the cluster interactions of all its sub-clusters $T\subset S$,

\begin{equation}
    \widehat{H}_B(\bm{\sigma_S}) = \langle H(\bm{\sigma})\;\rvert\;\bm{\sigma}_S\rangle - \sum_{T\subset S}\widehat{H}_C(\bm{\sigma_T})
    \label{eq:cluster-interaction}
\end{equation}

Equation \ref{eq:cluster-interaction} clarifies the meaning of a cluster interaction as the average contribution to the total energy coming solely from a single cluster $S\in B$ and none of its subclusters. Accordingly, we see that the terms in the cluster decomposition represent \emph{energetic interactions} among species occupying the sites of a cluster that are not captured by any lower-order interactions. Figure \ref{fig:interactions}a shows a visualization of the main effect, nearest neighbor pair, and a triplet cluster interactions as Cartesian tensors for a cluster decomposition of a \ch{CrCoNi} alloy.

In our presentation so far, we started with a representation of a cluster decomposition using a CE with a standard basis. However, since the cluster decomposition is basis agnostic, we can discard the concept of a basis altogether. In fact, in the fANOVA and related literature, a function is simply decomposed into its ANOVA representation by directly appealing to Equation \ref{eq:cluster-interaction} \cite{sobol_global_2001, hooker_generalized_2007}. This approach has been used in concurrent work \cite{lammert_cluster_2022}, presenting an axiomatic exposition of the cluster expansion and the cluster decomposition, which is in essence equivalent to the formalism of tensor product fANOVA decompositions \cite{jeon_characterization_2012}.

As the name \emph{analysis of variance} suggests, a cluster decomposition also comprises a decomposition of the variance of a Hamiltonian $H$ under the a-priori non-interacting product measure $P(\bm{\sigma}) = \prod_i \rho_i(\bm{\sigma}_i)$ \cite{hooker_generalized_2007},
\begin{align}
    \opp{Var}[H(\bm{\sigma})] &= \sum_{B\neq \emptyset}\sum_{S\in B}\opp{Var}[\widehat{H}_B(\bm{\sigma}_S)] \label{eq:anova}\\
    &= N\sum_{B\neq \emptyset} m_B||\widehat{H}_B||^2_2 \label{eq:var-weights}
\end{align}
where we used the fact that the variance of each cluster interaction is equal to its cluster weight $\opp{Var}[\widehat{H}_B(\bm{\sigma}_S)] = ||\widehat{H}_B||^2_2$. Further by using Equation \ref{eq:cluster-interaction}, we see that the effective cluster weights are the associated conditional variance with all lower order variances subtracted, i.e. the variance that can be attributed to a single cluster only and to none of its sub-clusters,
\begin{equation}
    \opp{Var}[\widehat{H}_B(\bm{\sigma_S})] = \opp{Var}[ H(\bm{\sigma})\;\rvert\;\bm{\sigma}_S] - \sum_{T\subset S}\opp{Var}[ H(\bm{\sigma})\;\rvert\;\bm{\sigma}_T]
\end{equation}

\begin{figure*}
    \centering
    \includegraphics[width=\textwidth]{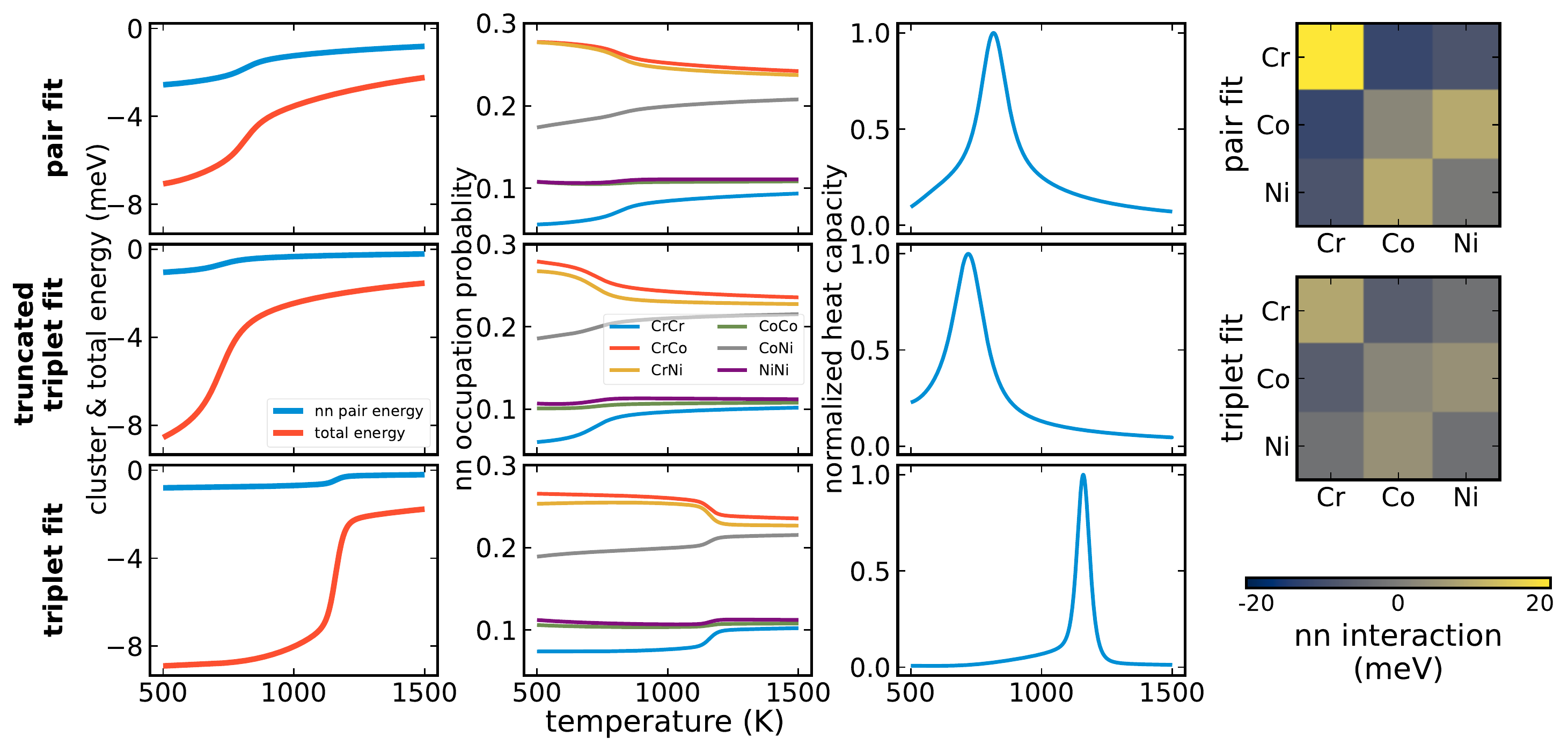}
    \caption{Nearest neighbor pair probabilities, cluster energies and total internal energy, normalized heat capacity, and nearest neighbor cluster interactions. The cluster energies of nearest neighbors are plotted with a solid blue curve, the total energy with a solid red curve, and the remaining cluster energies are plotted with dashed curves. Pair fit results (top), triplet fit results (bottom, solid), truncated fit (bottom, translucent/dot-dash).}
    \label{fig:crconi-sro}
\end{figure*}

Apart from providing a formal characterization of expansion terms, the cluster decomposition provides motivation and interpretations for the choice of regularization used when fitting. For example, Ridge regularization can be interpreted as setting an upper cutoff to the total variance. The use of Tikhonov regularization can be used as a way to more finely set variance cutoffs for specific cluster interaction terms. Recently proposed group-wise regularization \cite{barroso-luque_cluster_2022}, can be directly motivated as a judicious form to regularize cluster interactions $H_B$ by weighing coefficients with their permutation multiplicities $\widehat{m}_\beta$. Finally, estimation algorithms with hierarchical inclusion/exclusion of clusters \cite{van_de_walle_automating_2002, zarkevich_reliable_2004, leong_robust_2019, zhong_ensuremathell_0ensuremathell_2-norm_2022, barroso-luque_cluster_2022}, can be motivated by appealing to statistical concepts of \emph{hierarchically well-formulated} models \cite{peixoto_property_1990} that satisfy \emph{marginality constraints} \cite{mccullagh_generalized_2019}, or that abide by \emph{heredity principles} which satisfy either strong or weak hierarchy constraints \cite{hamada_analysis_1992, chipman_bayesian_1996}.

In addition, the cluster decomposition allows one to formally rank the importance of the contribution of each cluster interaction following the prescription of Sobol's sensitivity indices \cite{sobol_global_2001}. Accordingly, we define the effective cluster sensitivity index $\bar{\tau}_B$ as the fraction of the total variance of $H$ carried by the interactions of a cluster $S\in B$,
\begin{equation}
    \widehat{\tau}_B = \frac{\opp{Var}[\widehat{H}_B(\bm{\sigma}_S)]}{\opp{Var}[H(\bm{\sigma})]} 
\end{equation}

Similarly, we define the cluster sensitivity index $\tau_B$ as the normalized fraction of the total variance of $H(\bm{\sigma})$ contributed by the cluster interaction $\widehat{H}_B$ of all clusters $S\in B$, $\tau_B = m_B\widehat{\tau}_B$ per normalizing unit.  Cluster sensitivity indices provide a mathematically formal route for evaluating trends in the strength of interactions. Cluster sensitivity indices can be directly computed from a CE by using Equations \ref{eq:cluster-weight} for the cluster weights. Figure \ref{fig:interactions}b shows cluster sensitivity indices for the interactions of two fitted cluster decompositions of a \ch{CrCoNi} alloy.


As a basic example demonstrating practical use cases of the cluster decomposition, we fit two cluster expansions of a \ch{CrCoNi} medium entropy alloy. Our approach follows a recent study of the \ch{CrCoNi} alloy \cite{pei_statistics_2020} using a cluster expansion and Wang-Landau sampling \cite{wang_efficient_2001}. Following previous work, we fit two expansions \cite{pei_statistics_2020}: a less accurate expansion (in terms of cross-validation error) that includes pairs terms only (pair fit), and a more accurate expansion including pairs and triplets (triplet fit). We only reproduce previous results as an illustration and do not attempt to make any novel scientific claims about this particular alloy.

The energy contributions from interactions of specific species can be obtained by directly inspecting cluster interactions. Figure \ref{fig:interactions}a shows the main effect, pair, and triplet cluster interactions included in the triplet-fit cluster decomposition. We can readily determine which interactions are favorable (negative) and which are unfavorable (positive) based on the color map. The relative trend of the nearest-neighbor interactions obtained directly from the cluster decomposition agrees with previous results obtained via an ad-hoc, over-complete and less accurate nearest-neighbor pair model \cite{pei_statistics_2020}.

The interactions shown in Figure \ref{fig:interactions}a are of different orders of magnitude: the main effect contributions are of eV magnitude, and higher degree interactions are of meV magnitude. We can identify the most important cluster interactions, rank their importance, and compare different fits on rigorous grounds by using the corresponding cluster sensitivity indices as shown in Figure \ref{eq:cluster-interaction}b. In both fits, the first two pair interactions are the most important (largest sensitivity), with significant contributions coming from triplet interactions in the triplet fit.

As a further illustration of insights that can be obtained from the cluster decomposition, we computed nearest-neighbor pair short-range order, internal energy, and heat capacity of the \ch{CrCoNi} alloy from an equiatomic canonical Wang-Landau density of states using a 216 atom supercell. Figure \ref{fig:crconi-sro} shows the computed values for the pair fit and the triplet fit, as well as a truncated expansion including only the pair interactions from the triplet fit (triplet interactions removed). Comparing the nearest-neighbor pair energies and SRO results in Figure \ref{fig:crconi-sro} of the two decompositions that include only pair interactions, we observe that the overall SRO and total internal energy trends are set predominantly by the first and second nearest neighbor pair interactions (those with the highest cluster sensitivity from Figure \ref{fig:interactions}b). However, based on the triplet fit, we can conclude that triplet terms reduce the fraction of energy attributed to pair terms, tune the SRO values and raise the transition temperature. To delve deeper, one could inspect the triplet interaction values to better understand their role in tuning the ordering transition. These results agree with those reported previously \cite{pei_statistics_2020}, however, by using the cluster decomposition, we have shown how the results can be substantiated with a mathematically formal analysis.

We believe that substantially more insight, use cases, and parameter estimation methods beyond what has been presented here can be developed using the cluster decomposition and its formal statistical properties. The statistical literature is ripe with analysis techniques and methodology---such as log-density ANOVA models \cite{jeon_characterization_2012, gu_regression_2013-2} and sensitivity analysis \cite{sobol_global_2001, iooss_review_2015}---that can be directly leveraged in applications using parameterized lattice models. Several methods already exist in the statistics literature that can be used for direct estimation of cluster interactions and cluster indices in fully basis-agnostic manners \cite{sobol_global_2001, jeon_effective_2006, gu_regression_2013-2, iooss_review_2015}. Moreover, the formalism of the cluster decomposition is not limited to scalar functions of discrete degrees of freedom as presented here. In fact, a cluster decomposition can be obtained for any representation of scalar, vector, or tensor-valued function over a tensor product space domain by following the same prescription we have presented. Thus related expansions and generalizations \cite{drautz_spin-cluster_2004, thomas_exploration_2017, thomas_exploration_2017, drautz_atomic_2020} can be suitably recast as cluster decompositions and thus open the door to continued and significant developments based on rigorously established mathematical and statistical grounds.

An implementation of the cluster decomposition and all code used in this work is available at Ref \cite{barroso-luque_smol_2022-1}.

\begin{acknowledgments}
This work was primarily funded by the U.S. Department of Energy, Office of Science, Office of Basic Energy Sciences, Materials Sciences and Engineering Division under Contract No. DE-AC02-05-CH11231 (Materials Project program KC23MP). This research also used resources of the National Energy Research Scientific Computing Center (NERSC), a U.S. Department of Energy Office of Science User Facility located at Lawrence Berkeley National Laboratory, operated under Contract No. DE-AC02-05CH11231 using NERSC award BES-ERCAP0020531.
\end{acknowledgments}

\bibliography{refs.bib}

\appendix

\begin{widetext}
~
\nomenclature[01]{\(N\in \mathbb{N}_+\)}{{number of sites in a structure}}
\nomenclature[02]{\([N]=\{1, \ldots, N\}\)}{{index set for each of $N$ sites in a structure}}
\nomenclature[03]{\(\Omega_i\)}{{set of allowed species at site $i\in [N]$}}
\nomenclature[04]{\(\bm{\sigma}_i \in \Omega_i\)}{{occupancy variable for site $i\in[N]$}}
\nomenclature[05]{\(\bm{\sigma} = (\bm{\sigma}_1, \ldots, \bm{\sigma}_N)\)}{{occupancy string for a specific configuration of a structure with $N$ sites}}
\nomenclature[06]{\(S\subseteq[N]\)}{{cluster of sites}}
\nomenclature[07]{\(\bm{\sigma}_S = (\bm{\sigma}_i; \;\forall i\in S)\)}{{occupancy string of a cluster of sites $S$}}
\nomenclature[08]{\(B = \{\pi(S); \;\forall\pi\in\mathcal{G}\}\)}{{orbit of equivalent clusters under group $\mathcal{G}$; $D$ is also used as an orbit of equivalent clusters.}}
\nomenclature[09]{\(m_B = \lvert B\rvert/N\)}{{multiplicity of orbit $B$}}
\nomenclature[10]{\(\alpha = (\alpha_1, \ldots, \alpha_N)\)}{{multi-index, where each index $\alpha_i \in [0, \ldots, \lvert \Omega_i\rvert - 1]$; $\gamma$ is also used as a multi-index}}
\nomenclature[11]{\(\supp{\alpha} = \{i : \alpha_i\neq 0\}\)}{{support of a multi-index $\alpha$}}
\nomenclature[12]{\(\widehat{\alpha} = (\alpha_i : \alpha_i \neq 0)\)}{{reduced multi-index (only nonzero entries)}}
\nomenclature[13]{\(\beta = \{\pi(\alpha); \;\forall\pi\in\mathcal{G}\}\)}{{orbit of equivalent multi-indices under group $\mathcal{G}$}. $\eta$ is also used to represent orbits of multi-indices}
\nomenclature[14]{\(m_{\beta} = \lvert \beta\rvert/N\)}{{multiplicity of multi-index orbit $\beta$}}
\nomenclature[15]{\(\widehat{\beta} = \{\widehat{\alpha}; \;\forall \alpha\in\beta\}\)}{{orbit of equivalent reduced multi-indices under group $\mathcal{G}$}}
\nomenclature[16]{\(\widehat{m}_\beta = \lvert\widehat{\beta}\rvert\)}{{multiplicity of reduced multi-index orbit $\widehat{\beta}$}}
\nomenclature[17]{\(\phi_j : \Omega_i \to \mathbb{R}\)}{{univariate site basis function}}
\nomenclature[18]{\(H : \bigtimes_{i\in[N]}\Omega_i \to \mathbb{R}\)}{{function of configuration (Hamiltonian)}}
\nomenclature[19]{\(\rho_i : \Omega_i \to [0, 1]\)}{{a-priori probability measure for the occupancy of a site $i$}}
\nomenclature[21]{\(\Phi_\alpha : \bigtimes_{i\in[N]}\Omega_i \to \mathbb{R}\)}{{cluster basis function (expressed as a function over the full domain of all possible configurations of a structure). $\Psi_\alpha$ is also used as a cluster basis function}}
\nomenclature[22]{\(\Phi_{\widehat{\alpha}} : \bigtimes_{i\in S}\Omega_i \to \mathbb{R}\)}{{reduced cluster basis function (expressed as a function over the effective domain of the configurations of a cluster of sites $S = \supp{\alpha}$)}}
\nomenclature[23]{\(\Theta_\beta : \bigtimes_{i\in[N]}\Omega_i \to \mathbb{R}\)}{{correlation function}}
\nomenclature[24]{\(\widehat{\Theta}_\beta : \bigtimes_{i\in S}\Omega_i \to \mathbb{R}\)}{{reduced correlation function of a cluster of sites $S = \supp{\alpha}$ for $\alpha\in\beta$}}
\nomenclature[25]{\(H_B : \bigtimes_{i\in[N]}\Omega_i \to \mathbb{R}\)}{{mean cluster interaction of symmetrically equivalent clusters $S\in B$}}
\nomenclature[26]{\(\widehat{H}_B : \bigtimes_{i\in S}\Omega_i \to \mathbb{R}\)}{{cluster interaction of a cluster $S\in B$}}
\nomenclature[27]{\(\widehat{L}(B) =\{\widehat{\beta}:\;\opp{supp}(\alpha)\in B,\; \;\forall \alpha\in\beta\}\)}{{set of orbits of contracted multi-indices representing symmetrically distinct site function labelings over a cluster of sites $S\in B$}}

\markboth{}{} 
\printnomenclature[6cm]

\end{widetext}

\clearpage

\section{Change of basis matrices}

\begin{figure}[ht]
    \centering
    \includegraphics{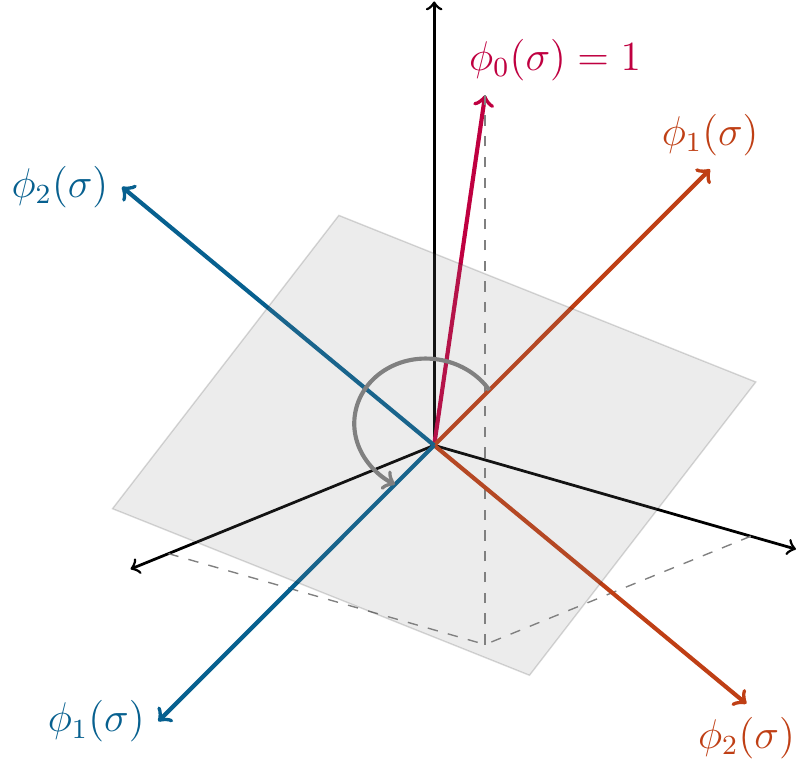}
    \caption{Two different choices of standard site basis sets for functions of the configuration for a ternary site space, and the rotation $\operatorname{R}$ relating them. Both basis sets by definition include the constant $\phi_0 = 1$ colored in red. Any arbitrary rotation about $\phi_0$ results in a standard site basis.}
    \label{fig:site-basis-rotations}
\end{figure}

For simplicity, let's consider a simple lattice system,\footnote{Extending to the general case with different site spaces is straightforward.} i.e. only one site space per lattice point. We start with a set of Fourier product basis functions constructed from a standard site basis $\{\phi_i, \;i=0, n -1\}$, written out as follows,
\begin{equation}
    \Phi_\alpha(\bm{\sigma}) = \prod_i^N\phi_{\alpha_i}
\end{equation}
Any another standard site basis $\{\psi_i\}$, must be related to $\{\phi_i\}$ by some rotation $\operatorname{R}$ orthogonal to $\phi_0$, i.e. $\psi_i = \operatorname{R}\phi_i$, as depicted in Figure \ref{fig:site-basis-rotations}. The resulting product basis functions can then be expressed as follows,
\begin{align}
    \Psi_\alpha(\bm{\sigma}) &= \prod_i^N\psi_{\alpha_i} \\
    &= \prod_i^NR\phi_{\alpha_i}
\end{align}

Now we can construct the change of basis matrix from $\Psi\rightarrow\Phi$ is $U_{\gamma\alpha} = \langle\Psi_\gamma, \Phi_\alpha\rangle$, starting from the following expression for the relation between the basis functions,
\begin{align}
    \Phi_\alpha(\bm{\sigma}) = \sum_{\gamma}\langle\Psi_{\gamma}, \Phi_{\alpha}\rangle\Psi_{\gamma}(\bm{\sigma})
\end{align}

By expressing the un-rotated basis $\Phi_\alpha$ in terms of the rotated basis $\Psi_\alpha$ we obtain the following expression for the change of basis matrix elements,
\begin{align}
    \langle\Psi_\gamma, \Phi_\alpha\rangle &=  \langle\prod_i^NR\phi_{\gamma_i},\prod_i^N\phi_{\alpha_i}\rangle \nonumber\\
    &= \prod_i^N\langle R\phi_{\gamma_i},\phi_{\alpha_i}\rangle \nonumber \\
    &= \left(\prod_i^NR_{\alpha_i,\gamma_i}\right)\delta_{\supp{\gamma}\supp{\alpha}}
\end{align}
where we used the fact that $\langle R\phi_{\gamma_i},\phi_{0}\rangle = \delta_{\gamma_i0}$, since by definition all non-constant functions must be orthogonal to $\phi_0$. We observe that the change of basis matrix is simply the product of elements of the site rotations matrix expressed in the $\{\phi_i\}$ basis for elements corresponding to product functions that act over the same cluster of sites $S$.

Since it is a change of basis matrix, $U_{\gamma,\alpha}$ must be orthogonal. But more importantly, the change of basis matrix $U_{\gamma\alpha}$, is block diagonal; where the blocks correspond to the product functions acting over the same set of site clusters $S$ identified by the support of their multi-indices ($\supp{\alpha}$). Furthermore, since $U_{\gamma,\alpha}$ is orthogonal, it follows that the blocks themselves are orthogonal. The blocks being orthogonal implies that for a given Hamiltonian $H$ the norm of all the expansion terms in a given block is left unchanged from a change of Fourier cluster basis,

\begin{align}
    \left\langle\left( \sum_{\gamma\ :\ \supp{\gamma} = S} J^{'}_{\gamma}\Psi_{\gamma}\right)^2\right\rangle &= \left\langle\left(\sum_{\alpha\ :\ \supp{\alpha} = S} J_{\alpha}\Phi_{\alpha}\right)^2\right\rangle \nonumber \\
    \sum_{\gamma\ :\ \supp{\gamma} = S} J^{'2}_{\gamma} &= \sum_{\alpha\ :\ \supp{\alpha} = S} J^2_{\alpha}
    \label{eq:norm-invariance}
\end{align}

The expression Equation \ref{eq:norm-invariance} above applies to any function of configuration, however, when dealing with a symmetrically invariant Hamiltonian, we can group the sums by site clusters $B$ and obtain the following invariance relation,
\begin{equation}
    \sum_{\eta\in L(B)} \widehat{m}_{\eta}J^{'2}_{\eta} = \sum_{\beta\in L(B)} \widehat{m}_\beta J^2_{\beta}
    \label{eq:weight-inv}
\end{equation}
where $L(B) = \{ \beta : \supp{\alpha} \in B \ \;\forall \alpha\in\beta\}$ are sets of function cluster orbits $\beta$ containing multi-indices $\alpha$ with symmetrically equivalent supports. Equation \ref{eq:weight-inv} is simply an expression that the cluster weights $||\widehat{H}_B||^2_2$ are invariant to the choice of basis.

\subsection{Properties of Fourier cluster and correlation functions}

First we show that Fourier cluster basis functions $\Phi_{\alpha}$ are normalized,
\begin{align}
    \langle \Phi_{\alpha}, \Phi_{\alpha} \rangle &= \left\langle \prod_{i=0}^{N - 1}\phi^{(i)}_{\alpha_i}(\bm{\sigma}_i), \prod_{i=0}^{N - 1}\phi^{(i)}_{\alpha_i}(\bm{\sigma}_i) \right\rangle \\
    &= \left\langle \prod_{i=0}^{N - 1}\phi^{(i)}_{\alpha_i}(\bm{\sigma}_i)\phi^{(i)}_{\alpha_i}(\bm{\sigma}_i) \right\rangle \\
    &= \prod_{i=0}^{N - 1}\langle \phi^{(i)}_{\alpha_i}, \phi^{(i)}_{\alpha_i}\rangle \\
    &= 1
\end{align}
Where we have used the fact that sums over configurations commute with products of site basis functions\footnote{Which means that site basis functions are \emph{uncorrelated} under a probabilistic interpretation}, and that standard site basis functions are orthonormal.

Now we show that Fourier basis functions are orthogonal following the same procedure,
\begin{align}
    \langle \Phi_{\alpha}, \Phi_{\gamma} \rangle &= \left\langle \prod_{i=0}^{N - 1}\phi^{(i)}_{\alpha_i}(\bm{\sigma}_i), \prod_{i=0}^{N - 1}\phi^{(i)}_{\gamma_i}(\bm{\sigma}_i) \right\rangle \\
    &= \left\langle \prod_{i=0}^{N - 1}\phi^{(i)}_{\alpha_i}(\bm{\sigma}_i)\phi^{(i)}_{\gamma_i}(\bm{\sigma}_i) \right\rangle \\
    &= \prod_{i=0}^{N - 1}\langle \phi^{(i)}_{\alpha_i}, \phi^{(i)}_{\gamma_i} \rangle \\
    &= \prod_{i=0}^{N - 1}\delta_{\alpha_i, \gamma_i} \\
    &= \delta_{\alpha, \gamma}
\end{align}

Fourier correlation functions can be shown to be orthogonal simply by expanding them in terms of Fourier product basis functions and using their orthonormality.

\begin{align}
    \langle \Theta_\beta, \Theta_\eta \rangle &=  \frac{1}{N^2} \left\langle\frac{1}{m_\beta} \sum_{\alpha\in\beta}\Phi_{\alpha}(\bm{\sigma}),  \frac{1}{m_\eta} \sum_{\gamma\in\eta}\Phi_{\eta}(\bm{\sigma}) \right\rangle \\
    &= \frac{1}{N^2m_\beta m_\eta}\sum_{\alpha\in\beta}\sum_{\gamma\in\eta}\langle \Phi_{\alpha}(\bm{\sigma}),\Phi_{\eta}(\bm{\sigma})\rangle \\
    &= \frac{1}{N^2m_\beta m_\eta}\sum_{\alpha\in\beta}\sum_{\gamma\in\eta}\delta_{\alpha, \eta} \\
    &= \frac{\delta_{\beta, \eta}}{N m_\beta}
\end{align}

For this, we used the fact that a multi-index $\alpha$ never appears in two different orbits $\beta \neq \gamma$. Note that correlation functions are not normalized with respect to the inner product used above.

Reduced correlation functions are similarly orthogonal but not normalized,
\begin{equation}
    \langle \widehat{\Theta}_\beta, \widehat{\Theta}_\eta \rangle = \frac{\delta_{\beta, \eta}}{\widehat{m}_\beta}
\end{equation}
Which can be derived by also expanding into cluster functions. Note that orbits of full multi-indices$\beta$ and contracted multi-indices $\widehat{\beta}$ can be used interchangeably considering the simple correspondence between all possible $\widehat{\beta}$ and all possible $\beta$ for any given symmetry group, i.e. simply take the set if contracted multi-indices $\widehat{\beta} = \{\widehat{\alpha};\; \;\forall \alpha\in\beta\}$.

\section{Properties of cluster interactions}

The proof of orthogonality of cluster interactions follows almost directly from the orthogonality of correlation functions,
\begin{align}
    \langle \widehat{H}_B, \widehat{H}_D \rangle
    &= \left\langle \sum_{\widehat{\beta}\in \widehat{L}(B)}\widehat{m}_\beta J_\beta\widehat{\Theta}_\beta,  \sum_{\widehat{\eta}\in \widehat{L}(D)}\widehat{m}_\eta J_{\eta}\widehat{\Theta}_\eta\right\rangle \\
    &= \sum_{\widehat{\beta}\in \widehat{L}(B)}\sum_{\widehat{\eta}\in \widehat{L}(D)}\widehat{m}_\beta \widehat{m}_\eta J_\beta J_\eta\langle\widehat{\Theta}_\beta, \widehat{\Theta}_\eta\rangle \\
    &= \sum_{\widehat{\beta}\in \widehat{L}(B)}\sum_{\widehat{\eta}\in \widehat{L}(D)}\widehat{m}_\beta \widehat{m}_\eta J_\beta J_\eta\frac{\delta_{\beta,\eta}}{\widehat{m}_\beta} \\
        &= \begin{cases}
        ||\widehat{H}_B||_2^2 \ \text{if} \ B = D \\
        0 \ \text{if} \ B \neq D
        \end{cases}
\end{align}

Showing that cluster interactions have mean zero follows directly from the derivation of orthogonality by setting $\widehat{H}_D = 1$.

Additionally, the irreducibly of a cluster interaction $\widehat{H}_B$, i.e that is orthogonal to any function $F_{\mathcal{D}}$ that can be expressed with reduced correlation functions $\Phi_\alpha$ with $\opp{supp}(\alpha)\in D$ for $D \in \mathcal{D}$ such that $B \notin\mathcal{D}$ also follows from the orthogonality of correlation functions. To show this we need to simply expand such a function in a Fourier correlation basis and use the fact that all basis functions will be orthogonal to those in the expansion of $\widehat{H}_B$.

Following a similar derivation, one can show that mean cluster interactions $H_B$ also have zero mean, orthogonal, and irreducible sets.

\section{Uniqueness of the cluster decomposition}

The proof of the uniqueness of a cluster decomposition is simple and follows established proofs for the uniqueness of the Sobol and the functional ANOVA decomposition \cite{sobol_global_2001, hooker_generalized_2007}. The proof is by contradiction, so we start by considering two different cluster decompositions for the same Hamiltonian $H$,

\begin{align}
    H(\bm{\sigma}) &= \sum_{B}\sum_{S\in B} \widehat{H}_B(\bm{\sigma}_S) \\
    H(\bm{\sigma}) &= \sum_{B}\sum_{S\in B} \widetilde{H}_B(\bm{\sigma}_S)
\end{align}

We can use the two expressions above as an expansion of a function everywhere zero, $F(\bm{\sigma}) = 0 \ \;\forall \bm{\sigma}$,
\begin{align}
    F(\bm{\sigma}) &= \left(\sum_{B}\sum_{S\in B}\widehat{H}_B(\bm{\sigma}_S) - \sum_{B} \sum_{S\in B}\widetilde{H}_B(\bm{\sigma}_S)\right) \\
    &= \sum_{B}\sum_{S\in B} \left(\widehat{H}_B(\bm{\sigma}_S) - \widetilde{H}_B(\bm{\sigma}_S)\right)
\end{align}

Now, if we consider the norm squared of expansion $F$ of the zero function,
\begin{flalign}
    &\langle F(\bm{\sigma}), F(\bm{\sigma}) \rangle = && \nonumber \\
    &= \sum_{B}\sum_{S\in B}\left\langle \left(\widehat{H}_B(\bm{\sigma}_S) - \widetilde{H}_B(\bm{\sigma}_S)\right), \left(\widehat{H}_D(\bm{\sigma}_S) - \widetilde{H}_D(\bm{\sigma}_S)\right) \right\rangle && \nonumber \\
    &= \sum_{B}\sum_{S\in B}\langle \widehat{H}^2_B(\bm{\sigma}_S)\rangle - \langle\widetilde{H}^2_B(\bm{\sigma}_S)\rangle = 0 && \label{eq:zero-fun}
\end{flalign}
Where we used the orthogonality properties of cluster interactions.

Finally, since the norm of a cluster the cluster interactions $\langle \widehat{H}^2_B(\bm{\sigma}_S)\rangle \ge 0$ and the norms of cluster interactions are invariant, each term in the sum in Equation \ref{eq:zero-fun} is equal to zero independently, which implies that the corresponding interactions must be equal (i.e. their differences are themselves zero functions),
\begin{equation}
    \widehat{H}_B(\bm{\sigma}_S) = \widetilde{H}_B(\bm{\sigma}_S)
\end{equation}
And so that the cluster decomposition of $H(\bm{\sigma})$ is unique.

\section{Custer variance decomposition}

The variance of a cluster expansion (under the a-priori product distribution), can be computed as follows,
\begin{align}
    \opp{Var}[H(\bm{\sigma})] &= \langle H^2 \rangle - \langle H \rangle^2 \\
    &= \sum_{\alpha} J^2_\alpha - J^2_{\bm{0}} = \sum_{\alpha\neq\bm{0}} J^2_{\alpha} \label{eq:full-var}
\end{align}
where $\bm{0}$ is the multi-index of all zeros, and we have used the orthonormality of Fourier cluster functions.

By grouping terms by multi-indices with the same support and subsequently, by symmetrically equivalent multi-indices, Equation \ref{eq:full-var} can be re-written as,
\begin{align}
    \opp{Var}[H(\bm{\sigma})] &= \sum_{B\neq \emptyset}\sum_{S\in B}\sum_{\widehat{\beta}\in \widehat{L}(B)} \widehat{m}_\beta J^2_\beta \label{eq:var-exp}\\
    &= N\sum_{B\neq \emptyset} m_B||\widehat{H}_B||^2_2
\end{align}
Where we identify the innermost sum in Equation \ref{eq:var-exp} as the norm of cluster interactions. Finally, also using Equation \ref{eq:full-var} it can be shown that the variance of a single cluster interaction is equal to its norm $\opp{Var}[\widehat{H}_B] = ||\widehat{H}_B||^2_2$.

\section{\ch{CrCoNi} alloy fit and computation details}

We fit two expansions using 500 training structures with up to 12 atoms per super-cell. The energy of the training structures was computed with density functional theory (DFT) using \textit{Vienna ab initio simulation package} (VASP) using the projector-augmented wave method\cite{kresse_efficiency_1996, kresse_ultrasoft_1999}, a plane-wave basis set with an energy cutoff of 520 eV, and a reciprocal space discretization of 200 \textit{k}-points per \AA. Electronic exchange-correlation effects are described using used the Perdew–Burke–Ernzerhof (PBE) generalized gradient approximation exchange-correlation functional \cite{perdew_generalized_1996}. All calculations were converged to $10^{-5}$ eV in total energy for electronic loops and 0.01 eV/\AA. DFT calculations were performed following the Materials Project \cite{jain_commentary_2013} \texttt{MetalRelaxSet} defined in the \texttt{pymatgen} Python package \cite{ong_python_2013}.

The cluster expansion fits were done using a mixed-integer quadratic problem (MIQP) formulation of a grouped $\ell_0$ pseudo-norm and $\ell_2$ norm regularization to obtain hierarchically constrained structured sparsity between cluster interactions \cite{zhong_ensuremathell_0ensuremathell_2-norm_2022, barroso-luque_cluster_2022}. This regularization ensures that \emph{strong hierarchy constraints} are respected in the resulting expansions \cite{peixoto_property_1990, hamada_analysis_1992, hamada_analysis_1992, chipman_bayesian_1996}. The regression optimization problem used is given as,
\begin{align}
    \min_{\bm{J}} ~& \bm{J}^{\top}\left(\bm{\Pi}^{\top}\bm{\Pi} + \lambda_1\bm{I}\right)\bm{J} - 2\bm{E}^{\top}\bm{\Pi}\bm{J} +  \lambda_0\sum_{B} z_{B} \label{eq:l2l0-miqp}\\
    &\begin{array}{r@{~}l@{}l@{\quad}l}
    \text{subject to} \quad Mz_{B}\bm{1} &\geq \bm{J}_B \\
    Mz_{B}\bm{1} &\geq -\bm{J}_{B} \\
    z_{B} &\in \{0,1\} \\
    z_B &\le z_D \;\ \;\forall D \text{ s.t. } D\sqsubset B
    \end{array} \nonumber
\end{align}
where $\bm{J}$ is a vector of all expansion coefficients, $\bm{J}_B$ are the coefficients corresponding to a single orbit $B$, $\bm{\Pi}$ is a matrix of correlation vectors of the training structures; $\bm{E}$ is a vector of DFT computed energies, $\bm{I}$ is the identity matrix, $\bm{1}$ are vectors of all ones, $\lambda_0, \lambda_1\in\mathbb{R}_+$ are hyper-parameters, $M\in\mathbb{R}_+$ is a fixed parameter, and $z_B$ are slack variables that describe whether a group of coefficients $\bm{J}_B$ associated with a single cluster interaction is zero or non-zero, i.e. active ($z_{B} \neq 0)$ or inactive ($z_{B} = 0)$. The notation $D\sqsubset B$ means that any cluster $T\in D$ is a subcluster $T\subset S$ of some cluster $S \in B$.

The final expansion fits are converged to a 5-fold root mean squared cross-validation of $12.8$ (triplet fit) and $14.9$ meV/atom (pair fit). The final fits include non-zero pair and triplet interactions with diameters up to $9 \AA$ and $4.3 \AA$, and non-zero pair interactions with diameters up to $7.5 \AA$ respectively, both based on a $2.49 \AA$  primitive lattice constant.

\end{document}